\documentclass[prd,aps,floats,preprintnumbers,preprint]{revtex4}
\usepackage{graphicx}
\usepackage{verbatim}
\usepackage{multirow}
\usepackage{longtable}

\newcommand{\be}{\begin{equation}}
\newcommand{\ee}{\end{equation}}
\newcommand{\bea}{\begin{eqnarray}}
\newcommand{\eea}{\end{eqnarray}}

\usepackage[colorinlistoftodos]{todonotes}

\begin{document}

\title{A new large CMB non-Gaussian anomaly and its alignment with cosmic structure}
\author{Antonio Enea Romano${}^{1,2,3}$, Daniel Cornejo ${}^{2}$, and Luis E. Campusano ${}^{2}$}

\affiliation{
${}^{1}$Department of Physics and CCTP, University of Crete, Heraklion 711 10, Greece \\
${}^{2}$Departamento de Astronom\'ia, Universidad de Chile, Casilla 36-D, Santiago, Chile\\
${}^{3}$Instituto de Fisica, Universidad de Antioquia, A.A.1226, Medellin, Colombia\\
}

\begin{abstract}
We provide evidence of the detection of a new  non-Gaussian anomaly in the cosmic microwave background (CMB) radiation which has larger statistical significance than the Cold Spot (CS) anomaly and comparable size.
This temperature anomaly is aligned with a huge large quasar group (HLQG),  and for this reason we call it HLQG anomaly.
There are different physical phenomena by which the HLQG could have produced the observed temperature anomaly, such as for example the Sunyaev Zeldovich (SZ), the integrated Sachs Wolfe (ISW) or the Rees Sciama (RS) effect. 
The goal of this paper it is not to explain the observed alignment in terms of these effects, but to show the shape and position of the HLQG anomaly, and estimate its statistical significance, i.e. the probability that it could be just the result of primordial Gaussian fluctuations.

We analyze the CMB Planck satellite temperature map of the region of sky corresponding to the angular position of the HLQG, 
and compute an inner and an outer temperature by averaging the CMB map over respectively the region subtended by the HLQG on the sky, and over a region surrounding it. It turns out that the inner region is warmer than the outer one, with a measured temperature difference of $\Delta T^{obs} \approx 43\mu K$. The temperature excess is  then compared with  the results of Montecarlo simulations of random Gaussian realizations of the CMB map, indicating with at least a $2.3\sigma$ confidence level, that the measured $\Delta T^{obs}$ cannot be attributed to primordial Gaussian fluctuations.

The HLQG anomaly has a higher statistical significance than the well known CS, and as such its detection provides a new important violation of Gaussianity.
Its angular extension in the longitudinal direction is about three times that of the CS, while its total angular area is comparable, due to its elongated shape compared to the CS.
Our results are stable under the choice of different simulations methods and different definitions of the inner and outer regions.  

\end{abstract}

\maketitle
\section{Introduction}
The  CMB radiation provides a unique window on the early Universe, and has allowed to
to test the standard cosmological model to a very high level of accuracy. The CMB temperature field is compatible with the inflation prediction that primordial curvature perturbations were Gaussian, but some  temperature anomalies such as for example the  Cold Spot (CS) have received a lot of attention, due to the low probability, about $1.85\%$ \cite{Cruz:2006fy}, that could be the result of primordial Gaussian perturbations.
In this paper we provide evidence of the detection of a new temperature anomaly of  comparable size which has larger statistical significance than the CS, i.e. a probability lower than $1.85\%$ to be the consequence of Gaussian primordial curvature perturbations.  
This temperature anomaly is aligned with a huge large quasar group (HLQG), which has a mean redshift of 1.27, characteristic comoving size of $\sim 500$ Mpc and was previously identified by analyzing galaxy catalogs. We call the observed temperature pattern HLQG anomaly, due to its alignment with the HLQG.
While there is a possibility that this alignment could be accidental, the detection of this temperature anomaly hints to the fact that the HLQG may be affecting the temperature of the CMB photons propagating though it.

There are different physical phenomena by which the HLQG could have produced the observed temperature anomaly, such as for example the Sunyaev Zeldovich (SZ), the integrated Sachs Wolfe (ISW) or the Rees Sciama (RS) effect. 
The goal of this paper it is not to explain the observed alignment in terms of these effects, but to show the shape and position of the HLQG anomaly, and estimate its statistical significance, i.e. the probability that it could be just the result of primordial Gaussian fluctuations, in which case the alignment would be accidental.

The analysis of the Data Release 7 (DR7) quasar catalog of the Sloan Digital Sky Survey has allowed to identify a huge large quasar group on Gpc scales (HLQG) \cite{2013MNRAS.429.2910C}. 
Some doubts have been raised about its physical existence and its size challenges the theoretical predictions of the $\Lambda CDM$ cosmological model for structure formation.

The HLQG was detected in the quasar catalog by 3-dimensional single-linkage hierarchical clustering  with an estimated significance of 3.8 $\sigma$ by comparison with a control sample of quasars within a redshift interval identical to the range covered by the LQG member quasars. The estimated mean quasar overdensity of the HLQG is $\delta_q = \delta \rho_q/\rho_q = 0.4$, which may be hypothesized to be representative of the average total density contrast.

A further corroboration of the HLQG has been provided \cite{2013MNRAS.429.2910C} using the distribution of Mg II absorbers in the same redshift interval of the member quasars. However, its reality has been questioned  \cite{Nadathur:2013mva},  based on the empirical fact that distribution of nearest-neighbor distances between quasars belonging to a certain subsample of the DR7QSO catalogue follows an approximately  Poisson point distribution, and on the assumption  that this can be used to generate simulated quasar samples. This assumption, while reasonable as a clever simplifying hypothesis for a first study of  the problem, is  only based on empirical evidence from a subsample, and would require to be supported independently by actual n-body simulations, that take into consideration that quasars are transient events within galaxies, to be fully trusted. Consequently  the statistical results based on it may not be a completely realistic estimation of the likelihood of these structures. This implies that their method underestimates the probability of formation of structures like the HLQG, since the real Universe galaxy and quasar spatial distributions are gravitationally structured, and are not just Poisson distributions.
On the other hand the detection of HLQG and similar structures has been confirmed by Park \cite{Park:2015ela}, while their statistical analysis is affected by the same unverified assumption  for the use of the Poisson distribution in simulations. Further evidence of the existence of HLQG was given by Hutsemekers \cite{huts} who investigated the alignment of quasar polarizations with the HLQG preferential axes.
Further support for the existence of these type of structures as compared to $\Lambda CDM$ simulations is provided by the analysis of the distribution of luminous red galaxy also in the Sloan Digital Sky Survey data release 7 \cite{Wiegand:2013xfa}, which is found to be  far from Poissonian using Minkowski functionals. It also shows significant deviations from the $\Lambda CDM$ mock catalogues on samples as large as $500h^{-1} Mpc$ (more than $3\sigma$) and slight deviations of around $2 \sigma$ on $700h^{-1} Mpc$.
 

For the search of anomalies in the CMB, we define an inner region subtending the HLQG and an outer region surrounding it, and then compute the temperature averaged over the two regions, and their difference.
We then perform a Montecarlo analysis based on random gaussian realizations of the CMB sky, and find  the statistical distribution of the simulated temperature difference. We  conclude that the possibility  that the observed temperature difference is simply the consequence of primordial gaussian fluctuations seeded by post-inflationary primordial curvature perturbations can be excluded with a confidence level of $\approx 2.5 \sigma$.
The statistical results are stable under the choice of different methods used to generate the simulated CMB maps, and different definitions of the inner region.

The detection of the HLQG anomaly can be considered the equivalent of the recently observed alignment of the cold spot with a very large void \cite{Szapudi:2014zha}, and as such provides further evidence of the importance of CMB observations as a tool for the study of large cosmic structures \cite{2014PhRvL.113b1301A} . 

\section{Analysis of the observed CMB map}

The method used in identifying the HLQG was based on a friend of a friend algorithm applied with a linkage scale of 100 Mpc, resulting in 73 member quasars spanning the 1.17-1.37 redshift range and a mean linkage lenght of 66 Mpc. Similarly to \cite{2013MNRAS.429.2910C} we associate a spherical volume to each quasar member with a comoving radius of 33 Mpc, and we define the inner region of the CMB temperature map as the projection over the celestial sphere of these 73 spheres. 

The outer region is instead defined as a strip enveloping the inner region, with an intermediate separating strip, to clearly separate them.
The purpose of these definitions is to define a measure of the difference between the temperature of the region of the CMB map which should be affected by the HLQG, and a surrounding one which should not.
This is motivated by the  theoretical expectation for the effects  of  inhomogeneities of the gravitational field on the CMB photons. In the case of supervoids an inner cold region surrounded by a hotter ring is the predicted Rees-Sciama effect, while for an overdensity like HLQG a similar but opposite pattern is expected.

In order to exclude artificial  selection effects related to the definition of the inner and outer region, we consider different prescriptions.
The $IN_1$ is the result of the projection on the sky map of the concave envelope containing all the spheres. 
The envelope is obtained by calculating the $\alpha$-shape, with an alpha value of 0.6 deg, over the set of circumferences of radius 33 Mpc  centered at each quasar.
The $IN_2$ region is obtained as a superposition of the projections of the 33 Mpc spheres on the sky map.

As can be appreciated in the table II and figure. (3), the different choices of the method used to define $IN_1,IN_2$ do not have any significant effect on the data analysis.

The outer region is defined as the region between two lines corresponding to the projection on the sky map of the two concave envelopes containing respectively spheres of radius $2*33$ Mpc and $4*33$ Mpc. centered at each quasar location. This definition leave a buffer between the inner and outer region in order to set a clear separation between the two.
The above definitions are represented in details in fig.(1-2).

For the calculation of the average temperatures we use the healpy implementation of HEALPix\cite{2005ApJ...622..759G} to analyze the Planck mission Data Release 1 Inpainted SMICA CMB map \cite{CMB_SMICA_MAP}, version 1.20.
We define the temperature difference according to
\be
\Delta T=T_{IN}-T_{OUT} \quad;\quad
T_{OUT}=\sum_{i\in OUT} T_i \quad;\quad
T_{IN}=\sum_{i\in IN} T_i  \\
\ee
where $T_{OUT}$ and $T_{IN}$ are the temperature averaged over the pixels contained in the outer and inner region respectively.
The same definition is applied for consistency of the analysis to the observed data and to the simulated maps. 

\begin{table*}
\begin{tabular}{|c|c|}
 \hline
Inner region &  $\Delta T^{obs}(\mu K)$ \\
\hline
 $IN_1$ &   $43.062$\\
 \hline
 $IN_2$ &   $45.343$
 \\
\hline
\end{tabular}
\caption{Observational data.  In the second column it is reported the  averaged temperature difference between the inner and outer region. All the averaged values are obtained from the Inpainted SMICA map \cite{CMB_SMICA_MAP}. The two rows correspond to the two different definitions of inner region.}
\label{Table1}
\end{table*}

\section{Simulations of CMB maps}

We use the SYNFAST routine of the HEALPIX \cite{2005ApJ...622..759G} package to generate 100 CMB map simulations. The mask which contain the inner and outer regions is then rotated randomly $10^6$ times for each map, giving in total $10^8$ random realizations.
The distribution of average temperatures is obtained from the average computed over all rotations and over all the different simulated maps, i.e. over the $10^8$ random realizations.
The random rotations are performed using an efficient random rotation matrix algorithm \cite{1980SJNA...17..403S}, to ensure the correct uniform probability  distribution of the randomly rotated regions over the 2-sphere .

In order to exclude the possibility that our results are affected by the procedure adopted to generate the simulated maps we use and compare the results of four different methods. Each method \textit{uses a different} power spectrum in SYNFAST.
\begin{itemize}
\item Method I: 
Simulations obtained using the power spectrum of the raw SMICA CMB \cite{CMB_SMICA_MAP} map calculated with the ANAFAST  subroutine\cite{2005ApJ...622..759G}. The raw map is masked for contaminated pixels using IMASK.
\item Method II:
Simulations obtained using the power spectrum of the Inpainted SMICA CMB map, calculated with ANAFAST.
\item Method III:
Simulations obtained using the best fit lensed power spectrum of the base+lowl+lowlike+highL data \cite{COSMO_PARAMS} from the Planck Cosmological Parameters Data Release 1.
\item Method IV:
Simulation obtained using the theoretical calculation (with CAMB) of the unlensed power spectrum corresponding to  the best fit parameters of the base+lowl+lowlike+highL data \cite{COSMO_PARAMS} from the Planck Cosmological Parameters Data Release 1.
\end{itemize}

\section{Interpretation}
The analysis we have performed shows that the CMB photons propagating through the region of sky subtended by HLGQ are on average warmer than the ones going through the surrounding region, as can be seen in Table I. The observed temperature difference can be excluded with a $2.3\sigma$ confidence level to be simply the result of primordial gaussian fluctuations, as shown in Table II and fig.(3). This means that the probability that the observed temperature pattern could be the result of a Gaussian primordial fluctuations is about $1\%$, which is less then the $1.85\%$ of the CS.

Assuming the RS effect is the cause of this temperature pattern this can be considered a rather convincing evidence that HLQG is  associated to a large scale spatial inhomogeneity of the gravitational potential.
Nevertheless since we have considered averaged temperatures, we can just infer information about the average gravitational potential. This implies that it is possible that HLQG is not  a single continuous extended overdensity, but a collection of two or more, since once the temperature is averaged the possible separation between substructures cannot be resolved.
Consequently, while our results provide evidence of an averaged large scale spatial inhomogeneity of the gravitational potential aligned with the HLQG, it does not exclude the possibility that the structure corresponding to the HLQG is composed by smaller disconnected substructures. The maximum longitudinal extension of the CMB anomaly we have found is about 35 degrees, making it much longer than the cold spot, while their total  area is comparable, due to the  elongated shape of the sky projection of the HLQG. The angular longitudinal size can be compared to the cold spot because it is located at about the same latitude on the opposite hemisphere.

To a certain extent the very definition of connected structure depends to the freedom of choice of the linking length used in the friend of friend algorithm used to detect these structures in galaxy catalogs \cite{Nadathur:2013mva}, so our results can be considered a  better method to assess the presence of the HLQG since is based on the detection of the actual effects of its gravitational potential, and as such is not subjected to the arbitrariness of the choice of the  linking length.

Since along the same line of sight there can be more than a quasar it is difficult to use the CMB alone to determine in details the substructure of the HLQG, but such an analysis could be performed in the future with improved CMB maps, in the attempt to find the imprints of every single quasar, for further confirmation they are actually connected.
Nevertheless our analysis provides for the first time a strong evidence that the HLQG has a spatially averaged density higher than the surrounding space, as it can be deduced from the CMB temperature averaged over the region of sky subtending it.

Finally it should be noted that the statistical analysis we have performed strictly speaking is giving the probability that the observed CMB temperature pattern aligned with the HLQG is not the consequence of primordial random gaussian fluctuations, but careful calculations are required to confirm that it can be entirely ascribed to the RS effect. Such an alignment itself could be accidental in principle, but the fact that also another CMB anomaly such as the cold spot is aligned with a large void \cite{Szapudi:2014zha}, while the theoretical calculations so far performed  predict a quite different temperature \cite{Nadathur:2014tfa,Tomita:2007db,Tomita:2009wz}, makes the accidental alignment a less probable, while still not impossible, explanation. It could be that some of the technical assumptions and approximations made in the theoretical calculations  of the RS effect, such as the gauge choice used for the cosmological perturbations or the choice of exact solutions of Einstein´s equations used for non-perturbative calculations, do not describe correctly the physical properties of these structures.

Since our results are based on averaging temperatures, and as a such are related to the spatial average of the gravitational potential, another important effect which may be necessary to take into account is the proper theoretical treatment of large scale inhomogeneities and the averaging of the Einstein´s  equations\cite{Buchert:2001sa,Buchert:2007ik,Buchert:1999pq,Buchert:1999er,Buchert:2002ij,Romano:2013kua,Romano:2011mx}, since back reaction effects could substantially modify the results of perturbation theory.
On the other side, a modification of General Relativity seems a less likely explanation given its success in many other experimental tests. While this goes beyond the scope of the present paper, it definitely deserves careful future investigation, in order to establish a consistent theoretical framework able to explain the observed alignment of CMB anomalies with large scale structures. 
The importance of a proper consideration of the effects of inhomogeneities in the analysis of cosmological and astrophysical data has been shown also in other contexts such as
interpretation of the baryon acoustic  oscillation (BAO) peak affected by relativistic structure formation \cite{Roukema:2014tta,Roukema:2015cwa} or the analysis of small-scale features \cite{Chiang:2015eza}.

\section{Conclusions}
We have analyzed the Planck CMB temperature map data in the region of the sky subtended by the HLQG, and have found a  temperature difference between that region and an outer one surrounding it, with a statistical significance  at approximately $2.3 \sigma$ confidence level.
The statistical significance is independent of the method used to define the inner region. The HLQG anomaly is in qualitative agreement with the theoretically expected effect of a large overdensity on the photons propagating though it, consisting in a warmer inner region contrasting with a colder outer one. This is the opposite of the CS, where a hot ring is surrounding the cold inner region. 
Despite this qualitative agreement the accidental alignment cannot be excluded until a careful comparison of theoretical predictions with observed data is performed.
On the other side the detection of the HLQG anomaly is in itself an observational fact, which can be attributed to primordial Gaussian perturbations with a probability lower than the CS. Whether its alignment with the HLQG can be explained or not in terms of know physical effects such as the ISW,  RS or SZ effects, the HLQG anomaly has a statistical significance higher than the CS and a comparable size, making it a new important violation of Gaussianity which deserves further study in the future.
Its importance is independent of our present ability of explaining its existence in terms of the effects of the HLQG on the CMB photons, in the same way the CS is an important CMB anomaly despite the absence of a conclusive theoretical explanation of its existence as the effect of a large void aligned with it.
In the future it will  be interesting to perform a systematic search for the alignment of other CMB anomalies with other very large cosmic structures detected through the analysis of galaxy catalogs, in order to determine if such alignments can be accidental or are in fact due to the effects of the propagation of the photons through the subtending structures.

\begin{table*}
Simulated Results: HLQG\\
\begin{tabular}{|c|c|c|c|c|c|}
\hline
Method & Zone & $\overline{\Delta T^{sim}}(\mu K)$ & $\sigma_{\Delta T(\mu K)}$ & p($\Delta T^{sim}>\Delta T^{obs}$) & $\frac{\Delta T^{obs}-\overline{\Delta T^{sim}}}{\sigma^{sim}}$\\
\hline
\multirow{2}{*}{I} & $IN_1$ &0.002& 17.768& 0.007& $2.423$\\
 & $IN_2$ & 0.003&17.977& 0.005& $2.522$\\
\hline
\multirow{2}{*}{II} & $IN_1$ & 0 & 18.134& 0.008&  $2.374$\\
 & $IN_2$ & 0 & 18.348& 0.006&  $2.471$\\
 \hline
 \multirow{2}{*}{III} & $IN_1$ & 0.002& 18.327& 0.009 & $2.349$\\
 & $IN_2$ & 0.001 & 18.574 & 0.007 & $2.441$\\
 \hline
 \multirow{2}{*}{IV} & $IN_1$ & 0 & 18.295 & 0.009 & 2.353 \\
 & $IN_2$ & 0 & 18.539 & 0.007 & 2.445\\
  \hline
\end{tabular}
\caption{Results for the $\Delta T_i$ on the Montecarlo simulations obtained with different methods. The columns correspond from left to right respectively to simulations method,  the inner region definition, the average inner outer region temperature difference, the standard deviation of the simulated  temperature difference frequency, the normalized  probability that the simulated temperature could be greater than the observed one, and the ratio between the observed temperature difference and the standard deviation obtained from simulations. The results are given for four different simulations methods, and for the two different definitions of the inner region. The upperscripts ${}^{obs}$ and ${}^{sim}$ stand respectively for observed and simulated. }
\label{Table2}
\end{table*}

\begin{figure*}[h]
\centering
\includegraphics[scale=0.7]{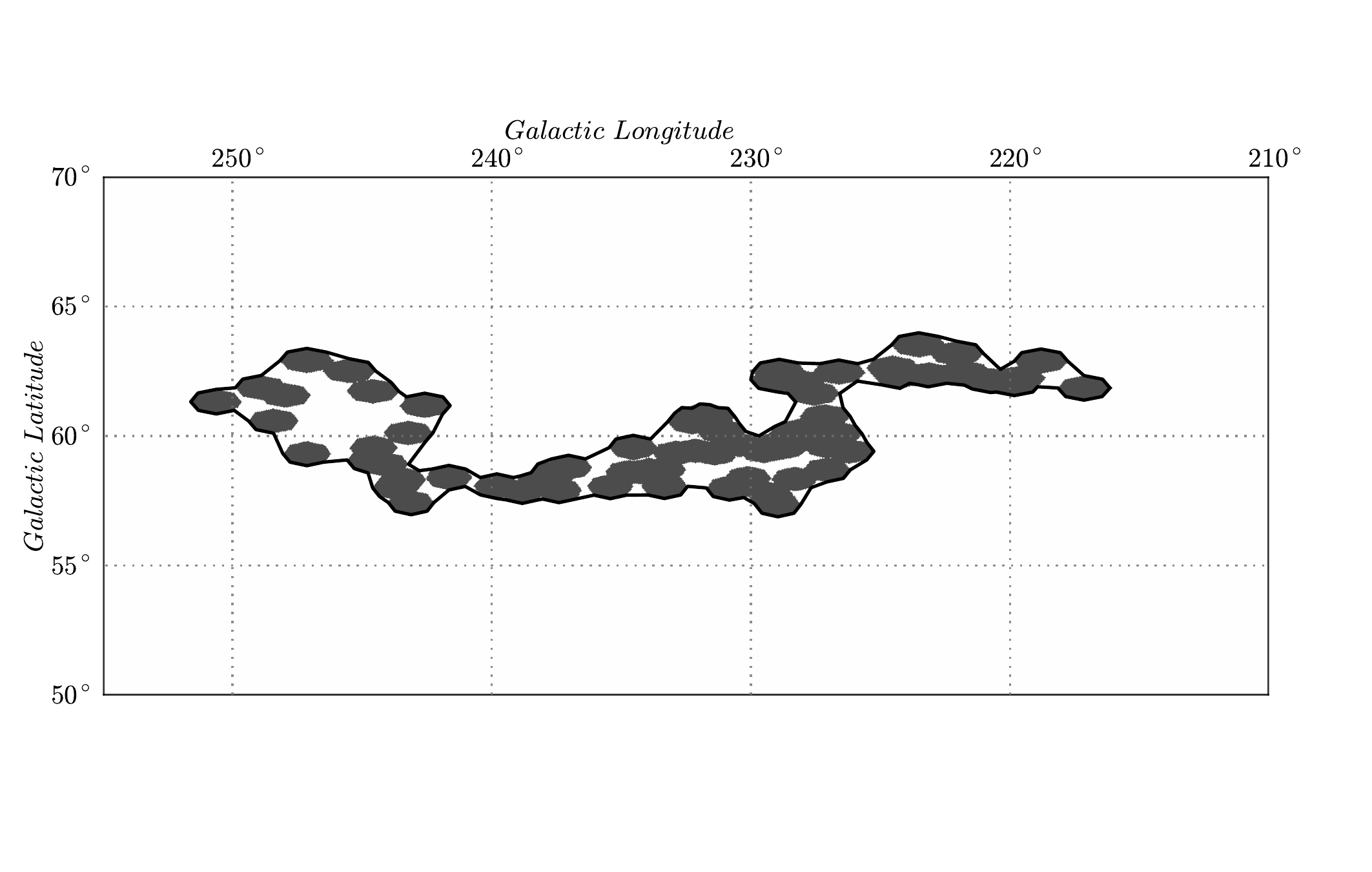}
\caption{
The black line is the frontier the inner region $IN_1$. The inner region $IN_2$ corresponds to the dark gray area, and is given by the projection on the sky of the individual quasars composing the HLQG. The map is a cartesian  projection of the galactic coordinates of the region of  sky subtending the HLQG.}
\label{fig:Figure1}
\end{figure*}
\clearpage


\begin{figure*}[h]
\centering
\includegraphics[scale=0.7]{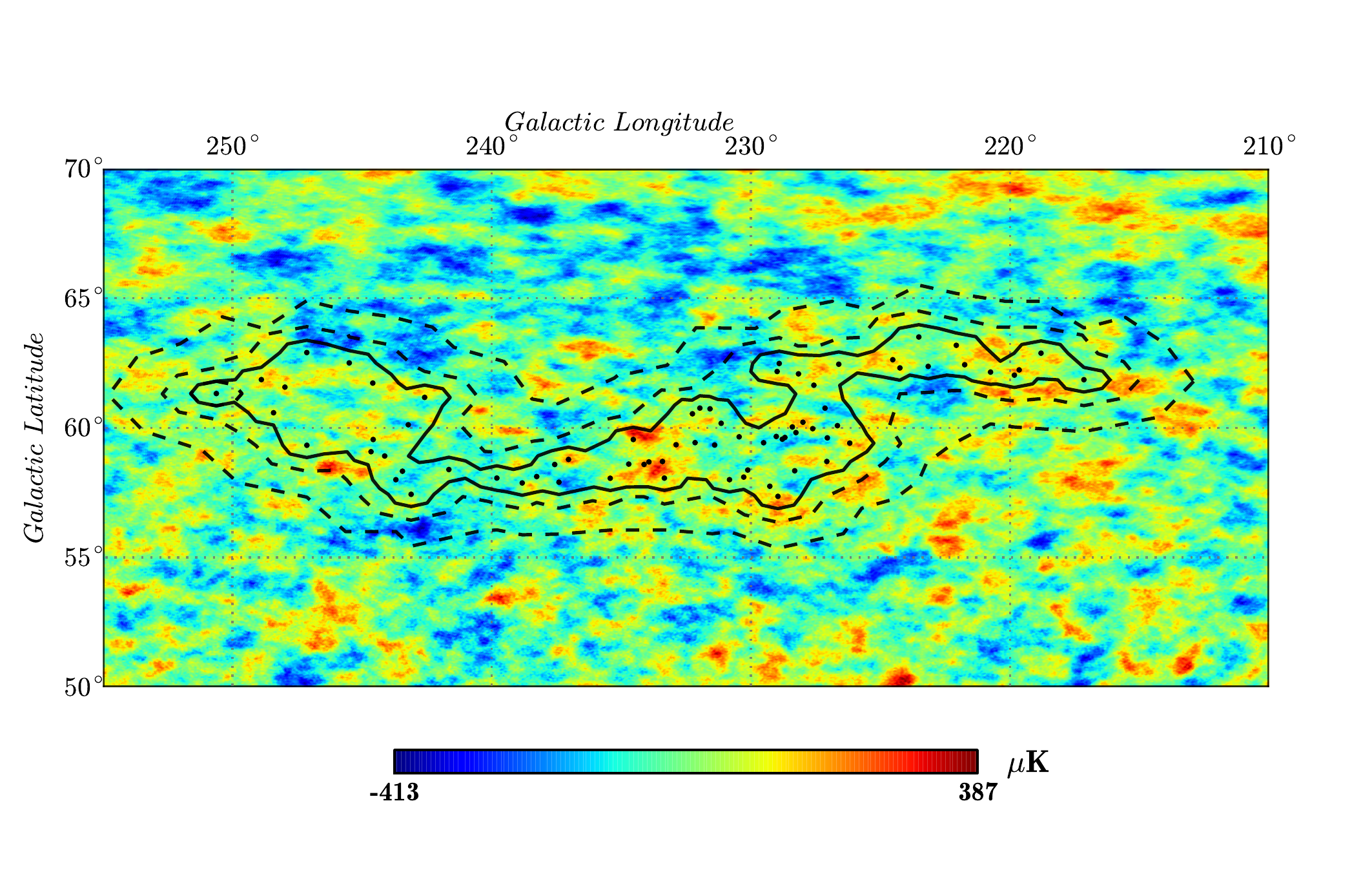}
\caption{
 The cartesian projection of the  SMICA CMB temperature anisotropy map is plotted for the region of sky of the HLQG. The two dashed black lines define the borders of the outer region. The solid black line define the inner region $IN_1$. The black dots correspond to the centers of the quasars.}
\label{fig:Figure2}
\end{figure*}
\clearpage



\begin{figure*}[h]
\centering
\includegraphics[scale=1]{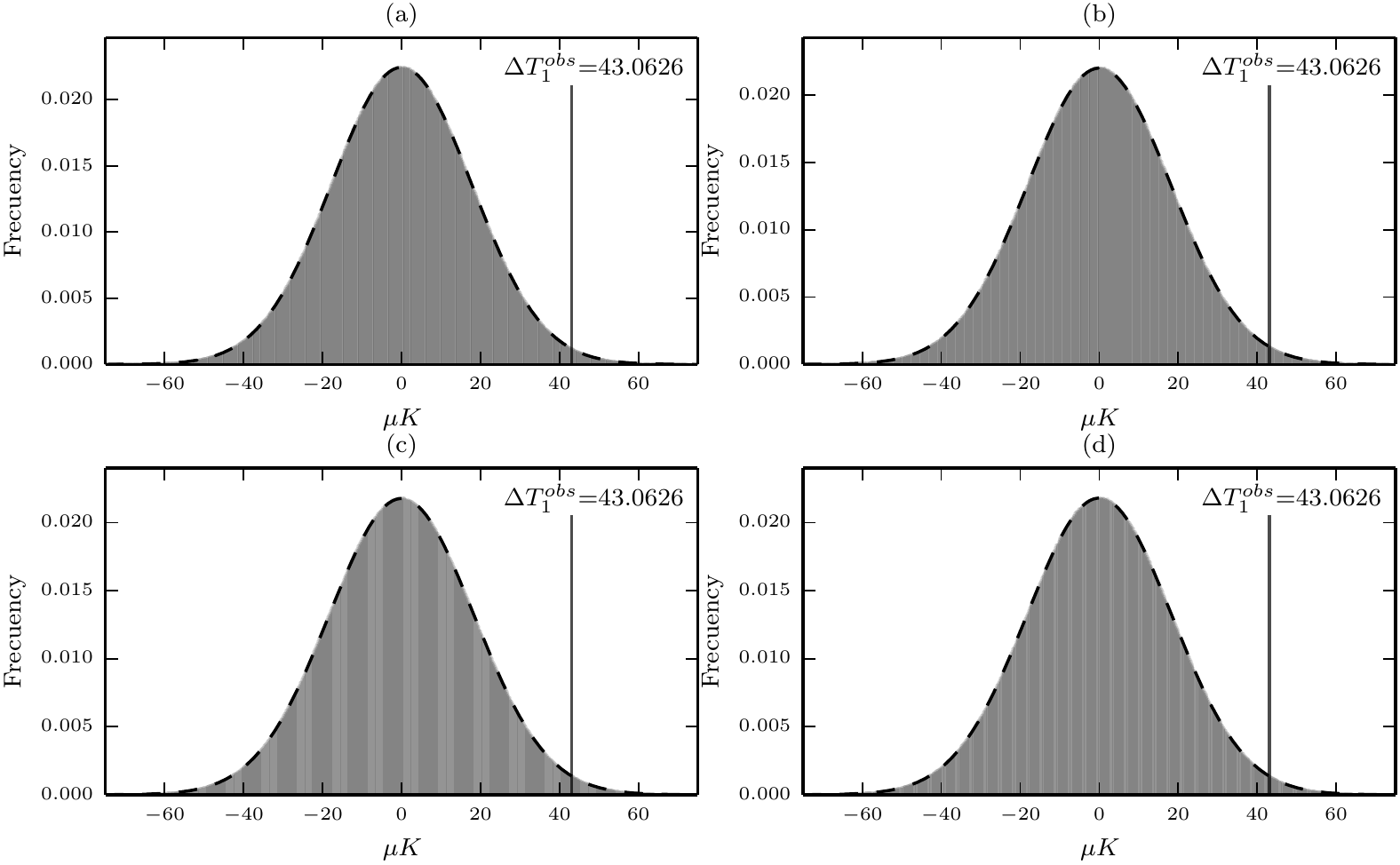}
\caption{
The normalized frequency of the simulated averaged temperature difference  $\Delta T_1$ is plotted for method I,II,III,IV respectively in panels  (a) , (b) , (c)  and (d). The inner region used is $IN_1$. The vertical line marks the observed temperature difference $\Delta T^{obs}_{1}$.}
\label{fig:Figure3}
\end{figure*}

\begin{figure*}[h]
\centering
\includegraphics[scale=1]{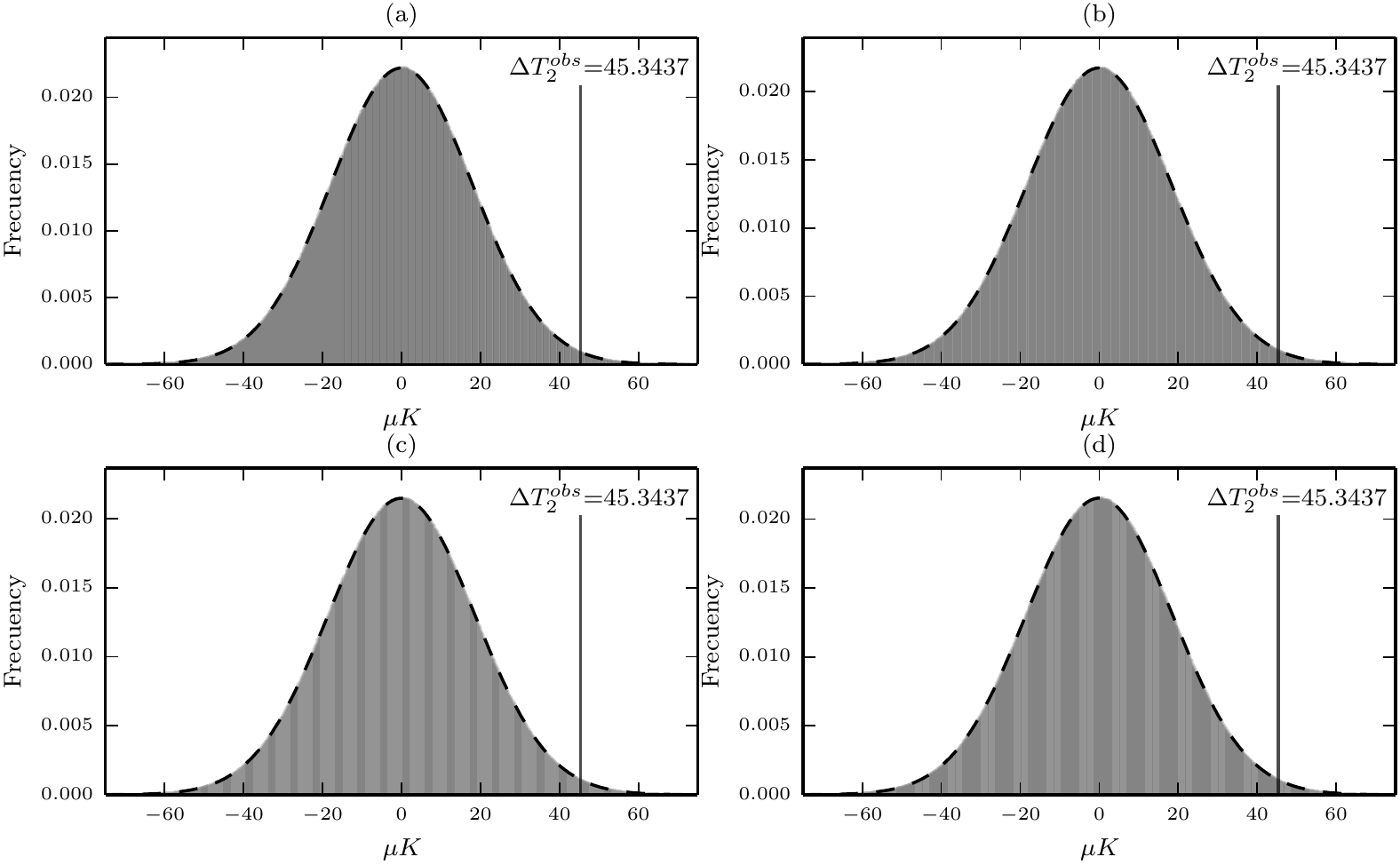}
\caption{
The normalized frequency of the simulated averaged temperature difference $\Delta T_2$ is plotted for method I,II,III,IV respectively in panels  (a) , (b) , (c)  and (d). The inner region used is $IN_2$. The vertical line marks the observed temperature difference $\Delta T^{obs}_{2}$.}
\label{fig:Figure4able}
\end{figure*}
\clearpage

\begin{acknowledgments}
AER thanks the Department of Physics and the Department of Astronomy of the Universidad de Chile for their kind hospitality. LEC thanks partial support from the Center of Excellence in Astrophysics and Associated Technologies (PFB-06). AER, DC and LEC, acknowledge partial support from CONICYT Anillo Project ACT-1122. We thank Hernan Pulgar, Srinivasan Raghunathan and Sebasti´an Pereira for their help during the course of the investigation.
This work was partially supported by the European Union (European Social Fund, ESF) and Greek national funds under the ``ARISTEIA II'' Action and the Dedicacion exclusica and Sostenibilidad programs at UDEA, the UDEA CODI
projects IN10219CE and 2015-4044..
\end{acknowledgments}

\bibliographystyle{h-physrev4}
\bibliography{lqgbib} 

{\color{red}

\end{document}